\documentclass[manuscript]{aastex}

\usepackage{graphicx}
\usepackage{epstopdf}

\shorttitle{Effect of CO on Haze Formation}
\shortauthors{H\"orst and Tolbert}

\begin{document}


\title{The Effect of Carbon Monoxide on Planetary Haze Formation}

\author{S.M. H\"orst\altaffilmark{1}, 
M.A Tolbert\altaffilmark{1,2}}
\altaffiltext{1}{Cooperative Institute for Research in Environmental Sciences, University of Colorado, Boulder, CO, USA}
\altaffiltext{2}{Department of Chemistry and Biochemistry, University of Colorado, Boulder, CO, USA}
\email{sarah.horst@colorado.edu}




\begin{abstract}
Organic haze plays a key role in many planetary processes ranging from influencing the radiation budget of an atmosphere to serving as a source of prebiotic molecules on the surface. Numerous experiments have investigated the aerosols produced by exposing mixtures of N$_{2}$/CH$_{4}$ to a variety of energy sources. However, many N$_{2}$/CH$_{4}$ atmospheres in both our solar system and extrasolar planetary systems also contain CO. We have conducted a series of atmosphere simulation experiments to investigate the effect of CO on formation and particle size of planetary haze analogues for a range of CO mixing ratios using two different energy sources, spark discharge and UV. We find that CO strongly affects both number density and particle size of the aerosols produced in our experiments and indicates that CO may play an important, previously unexplored, role in aerosol chemistry in planetary atmospheres.
\end{abstract}





\section{Introduction}

Atmospheric hazes, present in a range of solar system and extrasolar planetary atmospheres, play an important role in physical and chemical processes occurring in the atmosphere and for terrestrial planets, on the surface.  Haze particles affect the radiative balance of an atmosphere, may serve as condensation nuclei for clouds and rain, play a role in fluvial and aeolian processes \citep{Soderblom:2007, Burr:2006}, and affect the elemental budget of an atmosphere and surface. The effect on haze particles on the temperature structure of an atmosphere has implications for the habitability of a planet and an organic haze, such as seen in the atmosphere of Saturn's moon Titan, may potentially serve as the building blocks of life \citep{Horst:2012}.

While haze formation in N$_{2}$/CH$_{4}$ atmospheres has been extensively studied for decades in the laboratory through the production of Titan aerosol analogues or ``tholins'' (see \citet{Cable:2012}), the effect of other atmospheric constituents on the formation of haze in planetary atmospheres has not been well studied. Carbon monoxide (CO) is particularly interesting because it might serve as a source of oxygen for incorporation into photochemical aerosols affecting both their radiative and chemical properties and it is found in numerous atmospheres throughout the Universe. It is present in the hazy, reducing N$_{2}$/CH$_{4}$ atmospheres of Titan \citep{Lutz:1983, dekok:2007}, Pluto \citep{Greaves:2011,Lellouch:2011}, and Triton \citep{Lellouch:2010} and in the H$_{2}$ dominated atmospheres of the giant planets \citep{Beer:1975, Noll:1986, Marten:1993, Encrenaz:2004}. Recently CO and CH$_{4}$, in addition to handful of other molecules, have also been detected in the atmospheres of extrasolar planets (see e.g. \citet{Swain:2009, Madhusudhan:2009}) and haze layers have also been invoked to explain relatively featureless spectra of a number of exoplanets (see e.g. \citet{pont:2008}). It is therefore important to understand the effect of CO on the formation of planetary atmospheric hazes.

A few previous Titan atmosphere simulation experiments have included CO in their initial gas mixtures. \citet{Bernard:2003} and \citet{Coll:2003} focused on the effects of CO on the production of gas phase products. \citet{Tran:2008} focused on both gas and solid phase composition and observed the formation of ketones and carbonyls in their solid phase products. \citep{Horst:2012} reported the detection of amino acids and nucleotide bases in the solid products. However, none of these investigations reported the effect of CO on the size and number of haze particles produced, which are important parameters for the radiative effects of haze particles and for the total organic inventory found in haze particles. We present here an experimental investigation of the effect of CO on the formation of planetary atmospheric hazes including measurements of the size and number density of haze particles produced using a spark discharge source or UV photons to irradiate a range of mixtures of CH$_{4}$, CO, and N$_{2}$. 
	
\section{Materials and Experimental Methods}
\subsection{Haze Production Setup}

\begin{figure}
\plotone{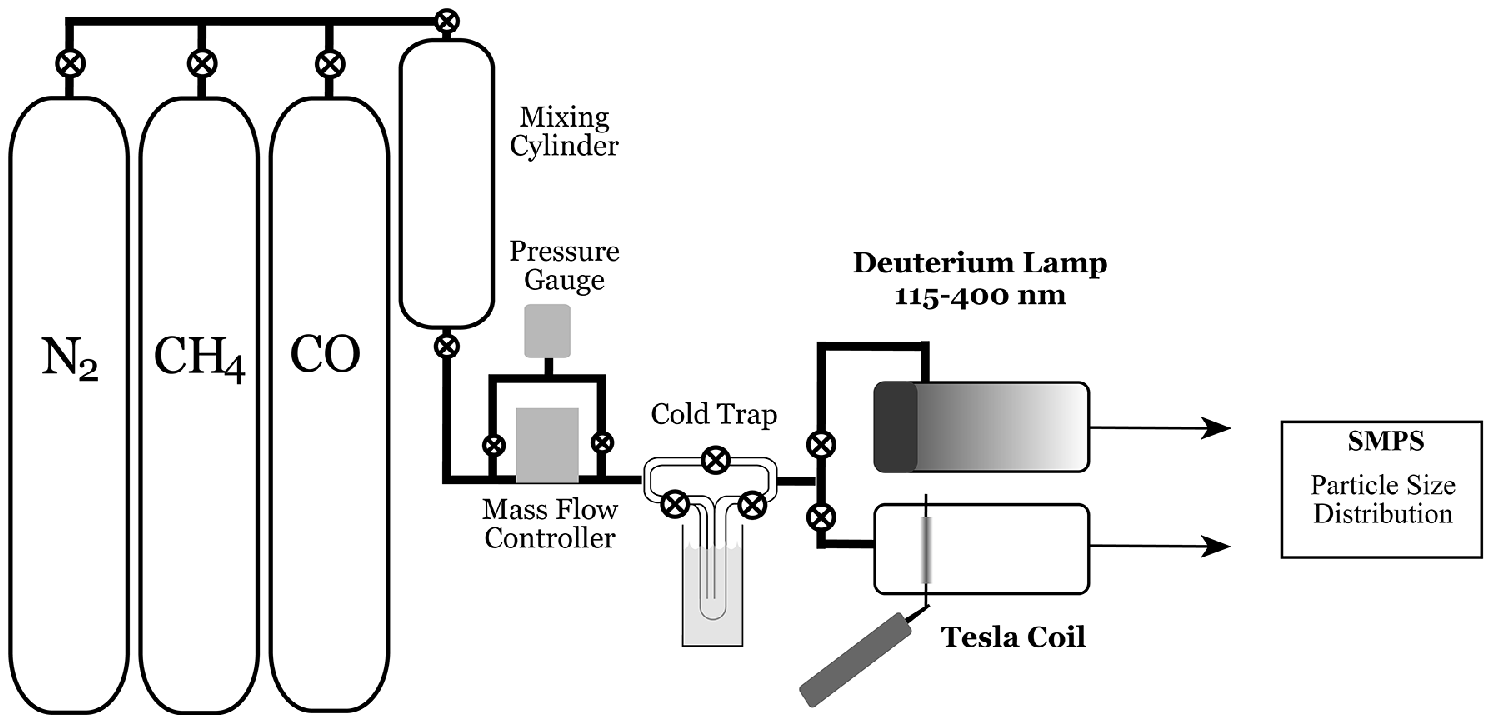}
\caption{Schematic of the experimental setup used for this work. N$_{2}$, CH$_{4}$, and CO mix overnight in the mixing cylinder. Gases flow through a cold trap held at -115$^{\circ}$C and into one of two reaction cells (UV or spark) where they are exposed to FUV photons from a deuterium lamp or the electric discharge produced by a tesla coil initiating chemical processes that lead to the formation of new gas phase products and particles. The particles are analyzed using a scanning mobility particle sizer (SMPS) to measure their size distribution. \label{fig:experiment}}
\end{figure}

Figure \ref{fig:experiment} shows a schematic of our experimental setup. Previous UV and spark experiments were performed using a similar setup by \citet{Trainer:2006,Trainer:2012,Trainer:2013, Horst:2013} and \citet{Trainer:2004, Trainer:2004b}, respectively. We introduced CO (99.999\% Airgas) in volume mixing ratios ranging from 50 ppm (the abundance in Titan's atmosphere \citep{dekok:2007}) to 5\% and CH$_{4}$ (99.99\% Airgas) in volume mixing ratios of 0.1\% and 2\% (see Table 1) into a stainless steel mixing chamber, then filled the mixing chamber to 600 PSI with N$_{2}$ (99.999\% Airgas). We allow the gases to mix for a minimum of 8 hours before running the experiment. The reactant gases continuously flowed through a cold trap, to remove trace impurities in the gases, before flowing through a glass reaction cell. We maintain a flow rate of 100 standard cubic centimeters per minute (sccm) using a mass flow controller (Mykrolis FC-2900).  The glass cold trap consists of two lines; one line is immersed in a slurry of 200 proof ethanol and liquid nitrogen, while the other line bypasses the cold trap entirely. For this work, the bypass line remained closed. The slurry remained at a temperature of $\sim$-115 $^{\circ}$C. The temperature of the gas line into the production cell was also monitored and was found to be unaffected by the use of the cold bath. We maintain the pressure in the reaction cell between 620 and 640 Torr (atmospheric pressure in Boulder, CO) at room temperature. We expose the reactant gases to one of two energy sources, spark discharge from a tesla coil or FUV photons, which initiate chemistry leading to particle formation. The experimental setup is the same for both energy sources until the gases reach the reaction cells. A tesla coil (Electro Technic Products) is connected to the spark reaction cell, while the UV reaction cell is connected to a deuterium lamp with a MgF$_{2}$ window (Hamamatsu L1835). 

Aerosol particle formation results from gas phase chemistry initiated by energy from the tesla coil or the deuterium lamp. Photons play a dominant role in the dissociation and ionization of chemical species that eventually lead to the formation of aerosols in Titan's atmosphere \citep{Lavvas:2011b}. It is therefore of paramount importance to investigate aerosol formation from photochemistry. The deuterium lamp we used for these experiments is a continuum source that produces photons from 115-400 nm (with major peaks near 121 and 160 nm). Although these photons are not sufficiently energetic to directly dissociate N$_{2}$ and CO, \citet{Trainer:2012} and \citet{Yoon:2013} demonstrated that nitrogen is participating in the chemistry in our reaction cell. Work is ongoing in our laboratory to understand the mechanism(s) responsible for the observed nitrogen incorporation. It seems likely that CO is dissociated through an analogous mechanism due to the similarity of their bonds. 

We use the electrical discharge because it is known to dissociate the triple bonds of CO and N$_{2}$ and is therefore an analog of the relatively energetic environment of upper atmospheres. However, we acknowledge that the resulting energy density is higher than the energy available in most planetary atmospheres to initiate chemistry. We use a tesla coil that can operate at a range of voltages. As described in \citet{Horst:2013}, we set the tesla coil to minimize the energy density while still producing sufficient aerosol using 2\% CH$_{4}$ in N$_{2}$ for our analytical techniques and used that setting for every experiment.

The flow exits the reaction cell and flows into a scanning mobility particle sizer (SMPS), which measures the distribution of particle sizes. The SMPS has three parts: an electrostatic classifier (TSI 3080), a differential mobility analyzer (DMA, TSI 3081), and a condensation particle counter (CPC, TSI 3775). The polydisperse aerosol first enters the DMA, where an electric field is applied to the flow of particles, which are then size selected based on their electrical mobility against the drag force provided by the sheath flow. Sheath flows of either 3 L/min or 10 L/min were used depending on the range of particle sizes produced in the experiment (covering $D_{m}$ ranging from 14.5 to 673 nm or 7.4 to 289 nm, respectively). Once size-selected, the particles enter the CPC where the number of particles is measured by light scattering. In this manner, we measure the number of particles as a function of their mobility diameter ($D_{m}$). Our standard flow rate of 100 sccm is determined by requirements of other instruments (see e.g. \citet{Horst:2013}) and is used here for consistency; however, the SMPS requires a higher flow rate. We therefore add an additional flow of N$_{2}$ after the particles exit the reaction chamber bringing the total flow rate to 260 sccm. The dilution caused by the additional flow of N$_{2}$ is accounted for during data analysis. 

Our \emph{in situ} analysis technique prevents the particles from being exposed to Earth's atmosphere and does not require sample collection, which could alter the observed particle sizes and number densities. However, real time analysis requires higher production rates. Additionally, the SMPS requires pressures at or near atmospheric pressure for operation. For those two reasons, we ran the experiments presented here at 620-640 Torr (Boulder, CO, atmospheric pressure, altitude $\sim$1600 m). This pressure is higher than the surface pressure on Pluto and Triton and in Titan's thermosphere where the chemical processes that result in formation of aerosol begin. Here we are interested only in comparing differences resulting from the addition of CO and from the choice of energy source at our standard experimental pressure. 

Our previous Titan simulation experiments have used 0.1\% CH$_{4}$ for UV experiments \citep{Trainer:2006, Trainer:2012, Trainer:2013, Horst:2013} and 2\% CH$_{4}$ for spark experiments \citep{Trainer:2004, Trainer:2004b, Horst:2013}. While 2\% CH$_{4}$ is analogous to Titan's atmosphere, the 0.1\% CH$_{4}$ is determined by experimental production constraints. Aerosol formation in our setup from the FUV lamp peaks near 0.1\% CH$_{4}$ due to optical depth in the reaction cell \citep{Trainer:2006, Horst:2013} and previous work has determined that the aerosol composition does not vary strongly with CH$_{4}$ concentration \citep{Trainer:2006}. Both for comparison purposes and to extend the range of planetary atmospheres where our results may shed light on aerosol formation, experiments were run for a range of CO concentrations using 0.1\% and 2\% CH$_{4}$. In Titan's atmosphere the CH$_{4}$ concentration has almost certainly varied over time, and since the CO abundance is tied both to CH$_{4}$ chemistry and the plumes of Enceladus \citep{Horst:2008}, the CO abundance has almost certainly varied over time as well. Measurement of absolute mixing ratios of CO and CH$_{4}$ in exoplanet atmospheres is still quite difficult (see e.g. the discussion in \citet{dekok:2013}) and further emphasizes the need to explore a range of CO and CH$_{4}$ concentrations.

\section{Particle Mass Loading and Size \label{sect:den}}

\begin{figure}
\plotone{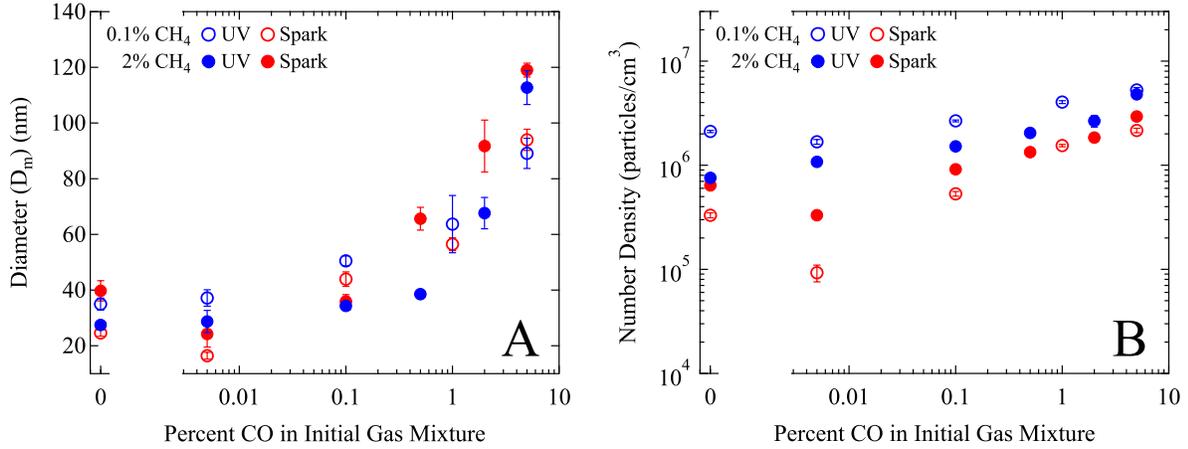}
\caption{Particle size (Panel A, mobility diameter ($D_{m}$), nm) and particle number density (Panel B, particles/cm$^{3}$) increase as a function of increasing CO in the initial gas mixture for both UV (blue) and spark (red) energy sources. This trend is observed for experiments using 0.1\% CH$_{4}$ (empty circles) and 2\% CH$_{4}$ (filled circles). Note that due to the small particle size, the full distribution for the 0.1\% CH$_{4}$, 50 ppm CO spark experiment could not be measured. Since only the tail end of the distribution was not measured, the number density was not strongly affected, but should be considered a lower limit. Error bars represent 1$\sigma$ error on multiple measurements. \label{fig:diameter}}
\end{figure}

\begin{figure}
\plotone{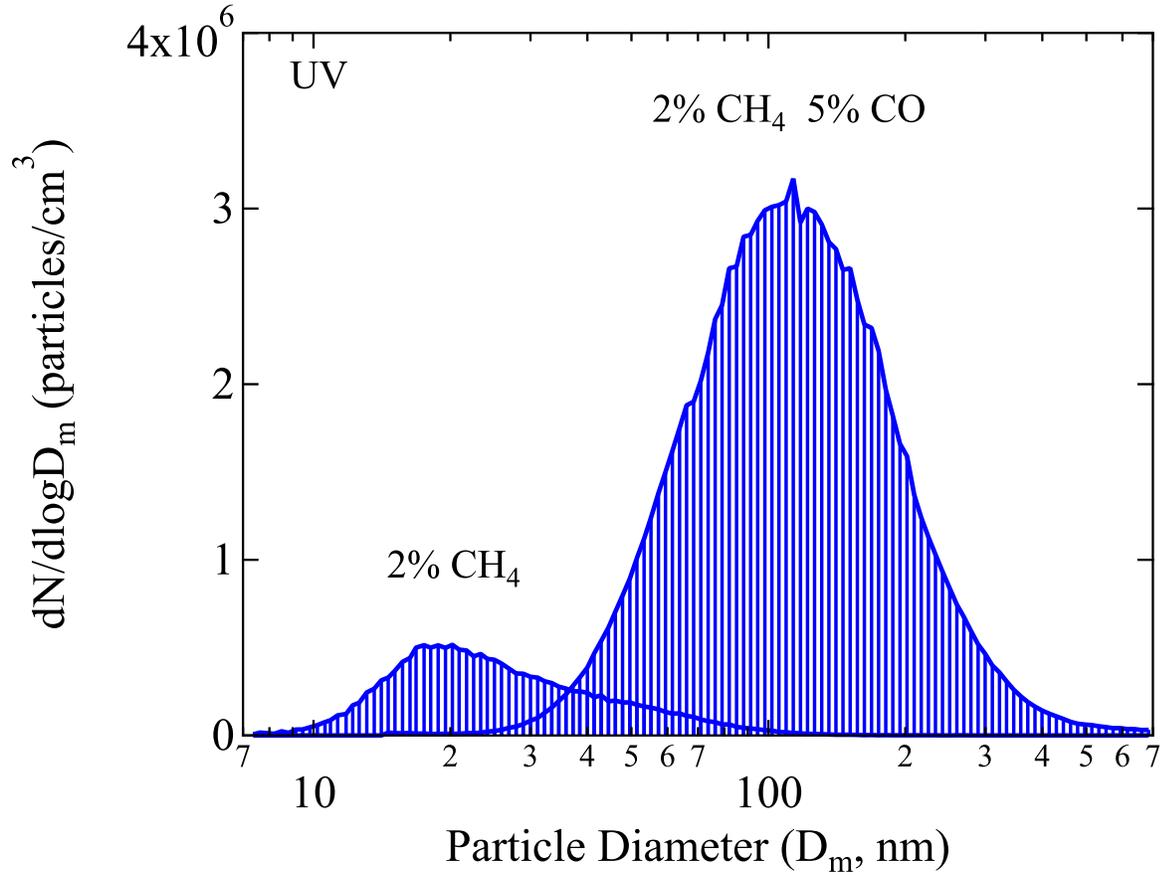}
\caption{Comparison of SMPS measurements of the size distributions of 2\% CH$_{4}$ and 2\% CH$_{4}$ 5\%CO UV samples. The data shown here are individual measurements and have not been corrected for dilution due to the additional N$_{2}$ flow. \label{fig:dist}}
\end{figure}

The SMPS measurements of particle size as a function of initial CO concentration are shown in Panel A of Figure \ref{fig:diameter}. For the spark experiments, the addition of 50 ppm of CO results in a decrease in particle diameter compared to experiments performed with no CO, while even the addition of 50 ppm CO results in an increase in particle size for the UV experiments.  In general, the particle diameter increases as a function of increasing CO starting at 50 ppm for both spark and UV energy sources regardless of methane concentration. Remarkably, the addition of 5\% CO results in the formation of particles with diameters 2-3 times larger than the experiments that did not include CO; particle distributions emphasizing this point are shown in Figure \ref{fig:dist}. While the strong influence of CO on particle diameter could potentially be attributed to the increase of carbon in the system, this behavior is not observed as a function of increasing CH$_{4}$ concentration in N$_{2}$/CH$_{4}$ in the same experiment \citep{Horst:2013}. For UV experiments, particle size decreases as a function of increasing CH$_{4}$ from 0.01\% to 10\% CH$_{4}$, while for the spark experiments the particle diameter increased until a peak at 2\% CH$_{4}$ and then decreased. 

The number density of particles, shown in Panel B of Figure \ref{fig:diameter}, exhibits behavior very similar to that of particle diameter; the number density increases as a function of increasing CO concentration for both energy sources at both CH$_{4}$ concentrations from 50 ppm CO to 5\% CO. As we observed with the N$_{2}$/CH${4}$ only experiments \citep{Horst:2013}, the UV experiments always produce more particles than the spark discharge experiments. For number density, the addition of 50 ppm CO results in decreases compared to no CO for both spark experiments (0.1\% CH$_{4}$ and 2\% CH$_{4}$) and a slight decrease for the 0.1\% CH$_{4}$ UV experiment. However, the addition of 50 ppm CO to the 2\% CH$_{4}$ UV experiment results in an increase in number density. This indicates that the presence of CO in Titan's atmosphere cannot be ignored in Titan atmosphere simulation experiments. 

The aerosol mass loading calculations are performed assuming that the density of the particle is 1 g/cm$^{3}$. An extensive discussion of the assumptions made in analysis of SMPS measurements, as well experimental determinations of tholin particle density for N$_{2}$/CH$_{4}$ experiments can be found in \citet{Horst:2013}. A variation of particle density is observed in both spark and UV experiments in the absence of CO. Density calculations require additional measurements not obtained for this work. However, the aerosol mass loading calculations are presented here, despite the assumption of density, so that the results may be compared to other works. Since both particle size and number density increase, aerosol mass loading also increases as a function of increasing CO concentration. The increase is most pronounced for the spark and 2\% CH$_{4}$ UV experiments. For the 2\% CH$_{4}$ UV experiment, an increase of more than 2 orders of magnitude is observed in the mass loading with a concentration of 5\% CO compared to the case where no CO is used. 

Taken together, the measurements of particle diameter and number density demonstrate that the addition of CO to N$_{2}$/CH$_{4}$ experiments results in more, larger particles. CO is therefore affecting both the formation and growth of aerosol particles in our experiments, even at relatively low concentrations. Further work, particularly on the composition of these particles, is necessary to fully understand the chemical mechanisms by which CO is affecting particle formation and growth. However, we have a few possible explanations for the observed behavior. 

First, previous tholin works have suggested that the presence of H$_{2}$  and H can decrease particle production. This explanation is often invoked to explain results of multiple plasma experiments which demonstrate that aerosol production first increases with increasing CH$_{4}$ concentration, reaches a peak, and then decreases. The decrease is attributed to an increase in the production of H$_{2}$ and H at high CH$_{4}$ concentrations \citep{Sciamma:2010, Horst:2013}. The addition of H$_{2}$ has also been shown to reduce aerosol formation in early Earth simulation experiments that photolyzed CH$_{4}$, CO$_{2}$, and N$_{2}$ \citep{dewitt:2009}. Perhaps oxygen, produced from CO in our experiments, is reacting with H$_{2}$ and H and removing them from the system. This could then lead to an aerosol formation by reducing one of the limiting factors. 

Second, as mentioned earlier, the presence of CO in the gas mixture does increase the total amount of carbon atoms present in the system. Based on our work looking at aerosol production as a function of CH$_{4}$ concentration, we know that simply increasing the amount of carbon in the system does not result in the formation of more aerosol for spark or UV experiments. However, the molecule that carries the carbon atom play a significant role. In the UV experiments, the decrease in aerosol production as a function of increasing CH$_{4}$ in the gas mixture has been attributed to the increase in optical depth at the wavelengths produced by our FUV lamp \citep{Trainer:2006, Horst:2013}. However, CO does not absorb these wavelengths and therefore may serve as source of carbon in the experiment without increasing the optical depth in the cell; thus more photons are available to drive aerosol chemistry. For spark experiments, increasing CO in the system increases the amount of carbon available without increasing the amount of hydrogen present in the system, which may also result in an increase of aerosol formation. 

Third, if the degree of oxygen participation in the chemistry is increasing as a function of increasing CO then the vapor pressures of the molecules produced may be lower, on average, than the molecules produced from N$_{2}$/CH$_{4}$ mixtures. This oxygen incorporation may shift the partitioning of gas and solid phase species toward the solid phase, which would result in the formation of more aerosol. 

The actual chemistry occurring in the reaction cell may be a combination of these three ideas or some other possibility. Future measurement of the aerosol and gas phase composition will provide insight into the partitioning of gas and solid phase species and allow us to assess the degree that oxygen is participating in the chemistry occurring in our experiment and the possible effect of CO on hydrogen chemistry. Isotopic labeling experiments will help determine the degree to which carbon in the aerosol originates from CO or CH$_{4}$.

\section{Conclusions}

We obtained \emph{in situ} particle size and number density measurements for tholins produced using CO concentrations from  50 ppm to 5\% and CH$_{4}$ concentrations of 0.1\% and 2\% and two different energy sources, spark discharge and UV as summarized in Table 1. For both energy sources and both CH$_{4}$ concentrations investigated, the particle size, number density and aerosol mass loading all increase as a function of increasing CO concentration above 50 ppm CO. The inclusion of CO has a dramatic effect on aerosol production, increasing the aerosol mass loading by orders of magnitude over the range of CO mixing ratios investigated. The fact that both the particle size and number density increase indicates that inclusion of CO increases both particle formation and growth. Intriguingly, the behavior as a function of CO mixing ratio is quite similar for both energy sources, in contrast to the behavior observed with only N$_{2}$/CH$_{4}$ gas mixtures where the production rate trends differ greatly with CH$_{4}$ concentrations based on energy source \citep{Horst:2013}.

Products of CO destruction may be decreasing the presence of H$_{2}$ and H in the reaction cell, species which are believed to inhibit aerosol formation, thus resulting in an increase in particle formation and growth. CO may effect aerosol formation by serving as an additional source of carbon, without affecting the optical depth at FUV wavelengths or introducing more hydrogen into the system. The increase in aerosol formation may also result from a shift in the partitioning between gas and solid phase species due to changes in vapor pressures of the molecules produced. However, further work is necessary to understand the effects of CO on the gas phase and particle phase composition before the role of CO in aerosol formation can be fully understood.

\acknowledgments
SMH is supported by NSF Astronomy and Astrophysics Postdoctoral Fellowship AST-1102827. This work was supported by NASA Planetary Atmospheres Grant NNX11AD82G.

\clearpage



\begin{figure}
\plotone{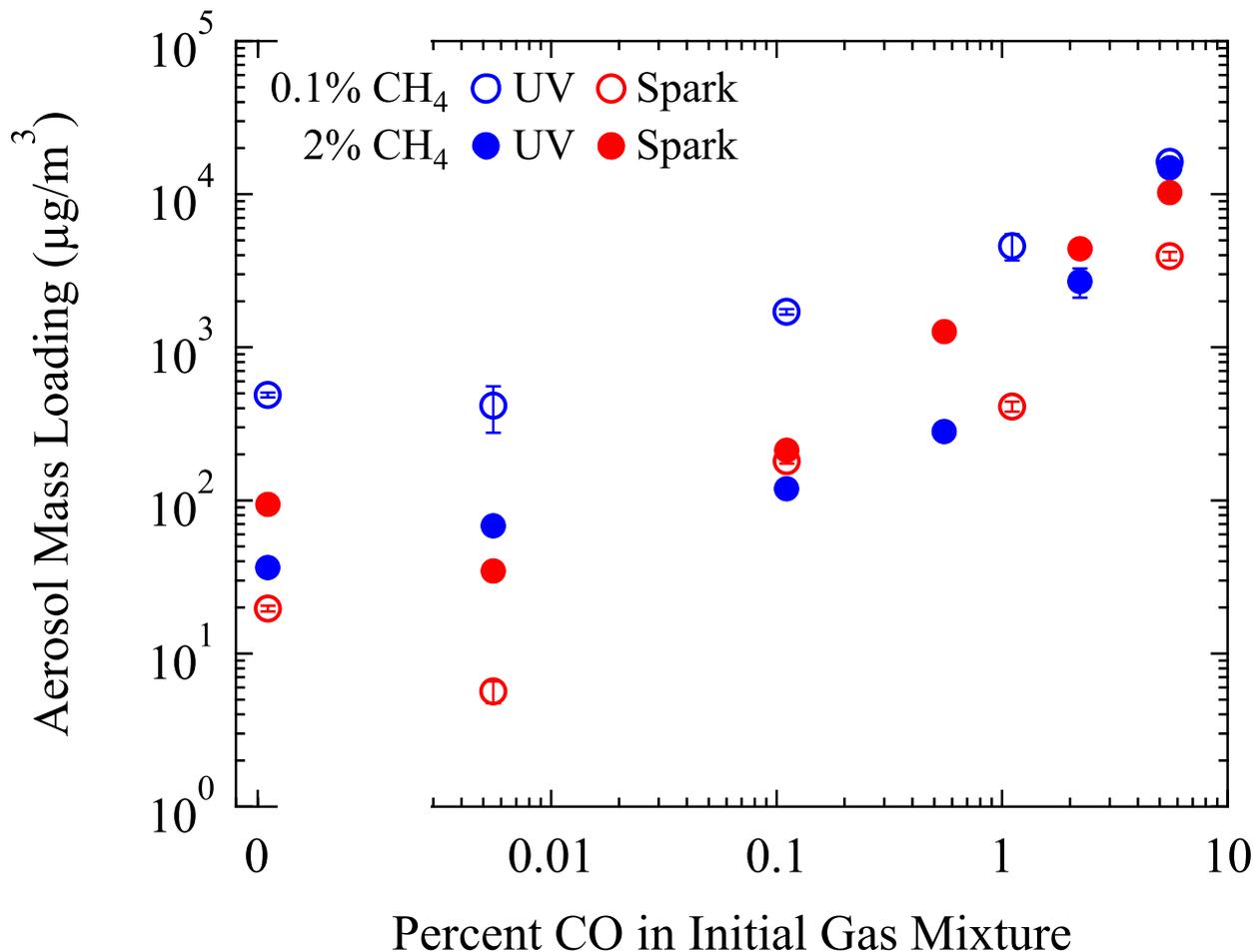}
\caption{Aerosol mass loading ($\mu$g/m$^{3}$), calculated by assuming a particle density of 1 g/cm$^{3}$, increases as a function of increasing CO in the initial gas mixture for both UV (blue) and spark (red) energy sources. This trend is observed for experiments using 0.1\% CH$_{4}$ (empty circles) and 2\% CH$_{4}$ (filled circles). Based on the particle size measurements and number density measurements shown in Figure \ref{fig:diameter}, the increase in aerosol loading is the result of both an increase in the number of particles and the formation of larger diameter particles. Note that due to the small particle size, the full distribution for the 0.1\% CH$_{4}$, 50 ppm CO spark experiment could not be measured. Since only the small particle tail end of the distribution was not measured, the mass loading was not strongly affected, but should be considered a lower limit. Error bars represent 1$\sigma$ error on multiple measurements.\label{fig:loading}}
\end{figure}

\clearpage

\renewcommand{\baselinestretch}{1}

\begin{deluxetable}{llll}
\tabletypesize{\small} 
\tablenum{1} 
\tablecolumns{3}
\tablewidth{0pt}
\tablecaption{Summary of Experiments Performed\label{Table:data}}
\tablehead{\colhead{CO}&\colhead{CH$_{4}$}&\colhead{CH$_{4}$}\\
\colhead{\%}&\colhead{0.1\%}&\colhead{2\%}}
\startdata
0&Spark, UV&Spark, UV\\
0.005&Spark, UV&Spark, UV\\
0.1&Spark, UV&Spark, UV\\
0.5&&Spark, UV\\
1&Spark, UV&\\
2&&Spark, UV\\
5&Spark, UV&Spark, UV\\
\hline
\enddata
\end{deluxetable}

\clearpage






\end{document}